\documentclass[12pt,draftcls,onecolumn,journal]{IEEEtran}
\ifCLASSINFOpdf
  \usepackage[pdftex]{graphicx}
\else
\fi

\usepackage{multirow}
\usepackage{subfig}
\usepackage{rotating} % sideways table
\usepackage{etoolbox}
\usepackage{float}

\usepackage{relsize}
\usepackage{caption}

\usepackage{textcomp}
\usepackage{mathptmx}
\usepackage{amsmath}
\usepackage{multirow}
\begin{document}
\title{Optimal Channel Sensing Strategy for Cognitive Radio Networks with Heavy-Tailed Idle Times}

\author{S.~Senthilmurugan,%~\IEEEmembership{Student~Member,~IEEE,}
        ~T.~G.~Venkatesh%~\IEEEmembership{Member,~IEEE}% <-this % stops a space
\thanks{The authors are with the Department
of Electrical Engineering, Indian Institute of Technology Madras, Chennai 600 036, India. e-mail:~amrita.senthil@gmail.com,~tgvenky@ee.iitm.ac.in.}% <-this % stops a space
%\thanks{Manuscript received April 19, 2005; revised January 11, 2007.}
}

\maketitle
\thispagestyle{plain}
\pagestyle{plain}

\begin{abstract}
%\boldmath
In Cognitive Radio Network (CRN), the secondary user (SU)  opportunistically access the wireless channels whenever they are free from the licensed / Primary User (PU). Even after occupying the channel, the SU has to sense the channel intermittently to detect reappearance of PU, so that it can stop its transmission and avoid interference to PU. Frequent channel sensing results in the degradation of SU's throughput whereas sparse sensing increases the interference experienced by the PU. Thus, optimal sensing interval policy plays a vital role in CRN. In the literature, optimal channel sensing strategy has been analyzed for the case when the ON-OFF time distributions of PU are exponential. However, the analysis of recent spectrum measurement traces reveals that PU exhibits heavy-tailed idle times which can be approximated well with Hyper-exponential distribution (HED). In our work, we deduce the structure of optimal sensing interval policy for channels with HED OFF times through Markov Decision Process (MDP). We  then use dynamic programming framework to derive sub-optimal sensing interval policies. A new Multishot sensing interval policy is proposed and it is compared with existing policies for its performance in terms of number of channel sensing and interference to PU.
\end{abstract}


% \fi
\IEEEpeerreviewmaketitle

\section{Introduction}

In recent years, usage of wireless devices such as smart phones, and laptops has grown exponentially. A major concern over this growth is that a large number of wireless devices are now trying to access  limited wireless spectrum. %The number of wireless technologies were very limited in early days and thus dedicated spectrum allocation for each technology was feasible.  
Further spectrum measurement campaigns have shown that the fixed spectrum assignment policy for wireless devices has resulted  in under utilization of the allotted bandwidth  \cite{Sadler2007}. Hence, to solve this problem and to have better spectrum utilization, researchers have proposed the technique of Cognitive Radio Network (CRN).
In CRN, the licensed bands are made available to unlicensed users, also called as Secondary Users (SUs)  whenever the licensed or Primary User (PU) are not using the spectrum.  
%Even after occupying the vacant channel, the SU has to sense the channel intermittently to avoid interference with PU. When a channel is reclaimed by PU, SU stops  transmission and goes to a sensing mode. 
In CRN, the channel sensing parameters of the SUs such as sensing time, sensing interval, and sensing accuracy have an impact on the performance of both secondary and primary network. Many works in literature \cite{Liang2008, Pei2011, Khosh2014, Shokri2015, Liang2011, pei2007} and references therein have studied the effect of PHY and MAC layer sensing parameters on the throughput of the secondary network. Most of these papers have assumed the ON and OFF time distribution of channel occupancy of the PU to be exponential.

%Earlier studies on optimal channel sensing mechanism assumes exponential PU ON-OFF time distribution .  

However, an in-depth analysis of spectrum measurement traces reveals that the idle times of ISM and GSM bands exhibit power law decay till some critical time after which it has exponential decay \cite{stabellini2010quantifying}, \cite{wellens2009empirical}. By power law decay, we mean that the log-log plot of probability density function $(p.d.f)$ of channel idle times, given by  $f_X(x) \propto x^{-a}$, will be a straight line with negative slope $-a$. The data sets with above behavior have been shown to be well modeled with Hyper-exponential distribution 
(HED) \cite{liu2012hyperexponential}, \cite{fitting}. Similarly, Sharma~\textit{et al.} \cite{sharma2011comprehensive} have simulated 802.11 WLAN clients-server model in OPNET simulator and  have observed that the channel idle times can be modeled using HED distribution.
The authors of \cite{huang2009optimization} and \cite{liu2014primary} have proposed an optimal SU sensing / transmission strategy  to maximize the throughput of SU with constraint on PU packet collision for  generalized as well as hyper-exponential PU idle time distributions. 

Many of the existing works have made the unrealistic assumption that SUs have full-duplex capability. A full duplex SU can transmit signal and detect the reappearance of PU at the same time.   However, the design of full-duplex system with acceptable PU detection probability is highly complex. Further, it increases the energy consumption of the SU. A promising use-case scenario of secondary network  is the wireless sensor network wherein it is not cost effective to deploy full-duplex SU. Considering the practical difficulties in implementing full-duplex SU, we look at the design of opportunistic secondary network that has half-duplex capability.
In a full-duplex system the SU can stop its transmission as soon as the PU is detected. Thus the interference to PU can be kept minimal (maximum of one PU packet as a result of packet header corruption). However in the case of  half-duplex system the  interference of SU with PU is in general more than that of full-duplex case and can be as large as the inter-sensing duration. Thus the problem of finding optimal sensing strategy becomes even more crucial in the half-duplex system.

The major contribution of our work which differs from the existing literature on channel sensing strategies \cite{pei2007}, \cite{huang2009optimization}, \cite{liu2014primary} in CRN are as follows: In contrast to \cite{pei2007}, we model PU OFF times as HED which is more realistic. Secondly, we have designed an optimal channel sensing interval framework considering half-duplex SUs whereas \cite{huang2009optimization}, \cite{liu2014primary} have assumed full-duplex SUs. Finally, we frame optimization problem that tries to minimise both the cost for number of channel sensing and the cost of interference to PU by choosing optimal channel sensing intervals. We have used dynamic programming to derive the optimal channel sensing interval policy in our work. Interest readers can look into  \cite{Shabara2015, Lee2008} on applying dynamic programming to other optimization problems in CRN.

The important insight of our work is that the constant periodic sensing policy is not an optimal solution for non-exponential PU OFF times. We have proved the above point by deducing the structure of optimal solution using Markov Decision Process (MDP). We further suggest a new sub-optimal policy called ''Multishot sensing interval policy" which outperforms existing sub-optimal channel sensing policies in the literature \cite{azad2011optimal}. 

The rest of the paper is organized as follows: A brief overview of system model is given in section~\ref{sec:sys_model}. In section~\ref{sec:opt_formulation}, we formulate the optimization  problem which balances the number of SU channel sensing and interference to PU. The structure of optimal solution is derived using MDP in section~\ref{sec:MDP}. In section~\ref{sec:suboptimal}, the different sub-optimal channel sensing interval policies are proposed . Section~\ref{sec:simulation} compares the performance of sub-optimal policies under various channel traffic conditions through simulation.
Section \ref{sec:SensingParams} studies the effect channel sensing parameters. Finally, we conclude the paper in section~\ref{sec:conclusion}.

\section{System Model}
\label{sec:sys_model}

\begin{figure}  
\begin{center}  
\includegraphics[scale=0.16]{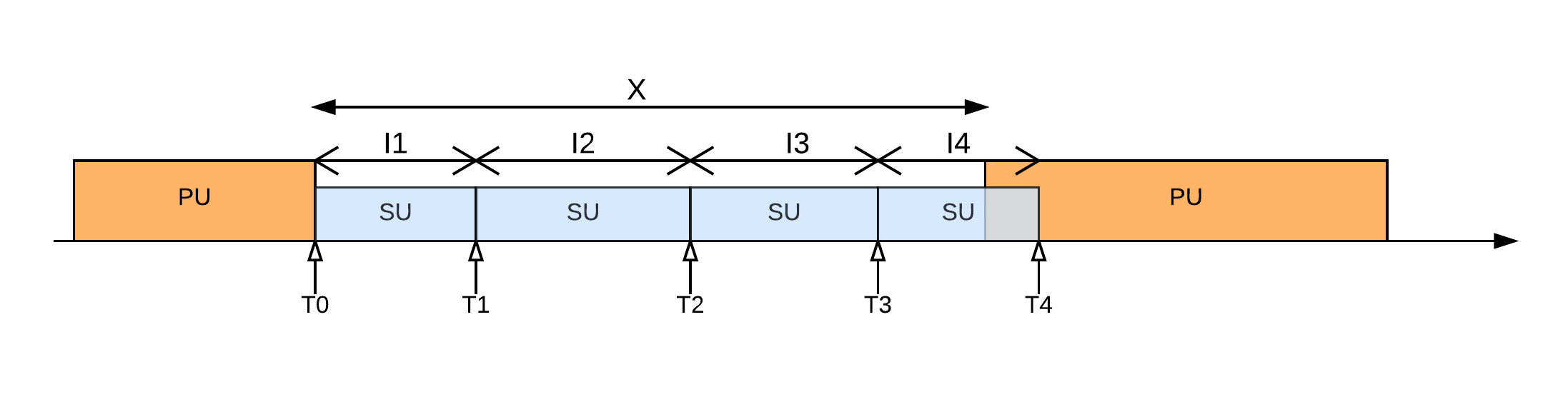}  
\caption{\small \sl Channel sensing strategy followed by SU to detect the presence of PU\label{fig:activity}}  
\end{center}  
\end{figure}

We consider CRN having half-duplex SUs. Following the studies on spectrum measurement traces, we model the OFF time distribution of PU to be heavy tailed. %The PU's channel occupancy pattern is statistically modeled as ON-OFF process. Following the recent measurement campaigns, the channel idle times are assumed to exhibit heavy tail distributions.  
As mentioned earlier, the heavy-tailed idle times (OFF times) of PU are well-modeled as K-phase HED distribution, % with parameters $(p_i)_{i=1}^{K}$ and $(\lambda_i)_{i=1}^{K}$ as
\begin{equation}
f_X (x) = \sum\limits_{i=1}^{K} p_i \lambda_i e^{-\lambda_i x}. \label{eq:fx}
\end{equation}
where $p_i$'s are the phase probabilities such that $\sum_{i=1}^K p_i = 1$, and $\lambda_i$'s are the rates of mixture of exponential distribution \cite{stabellini2010quantifying}. \footnote{ The random variable $X$ is said to follow HED if X is,  with probability $p_i$, exponentially distributed with parameter $\lambda_i$ for $i=1,2...,K$. }
The realistic spectrum measurement traces can be used to estimate the parameters $p_i$'s and $\lambda_i$'s of HED as shown in \cite{stabellini2010quantifying}.
Suppose the SU sense the channel to be free from the primary user and occupies it. Let $T_0$ be the time at which SU occupies the channel.  In order to avoid interference to PU, the half-duplex SU has to limit its transmission duration and intermittently sense the channel for the reappearance of PU. Let the sensing instants of SU fall at time instants $T_1$, $T_2$,...,$T_{N}$ where the index $N$ denotes the sensing instant at which SU detects the presence of PU. We denote the channel sensing intervals $(T_1 - T_0)$, $(T_2-T_1)$,..., $(T_{N}-T_{N-1})$ by $I_1$, $I_2$,..., $I_N$, respectively as indicated in Fig.~\ref{fig:activity}. Our formulation of the optimal sensing interval policy minimizes the number of sensing as well as the cost involved in the interference to PU.

We assume that SU immediately access the channel when PU goes from ON state to OFF state. We have also studied the effect of delayed occupancy of the channel in section~\ref{sec:SensingParams}. 
The residual PU OFF time at $j^{th}$ sensing instant, $X_j$, also has HED distribution with same $(\lambda_i)_{i=1}^{K}$  but with different $p_i$'s.  For example, the phase probabilities of residual PU OFF time $X_1$ at first sensing instant $T_{1}$, denoted by $p_1^1$, $p_2^1$,..., $p_K^1$, can be calculated as follows:
\begin{equation}
\begin{split}
Pr(X_1 > x) &= Pr(X > I_1 + x | X > I_1) = \frac{\sum\limits_{i=1}^{K} p_i e^{-\lambda_i (I_1+x)}}{\sum\limits_{k=1}^{K} p_k e^{-\lambda_k I_1}}
= \sum\limits_{i=1}^{K} p_i^1 e^{-\lambda_i x},
\end{split}
\end{equation}
where $p_i^1$ is given as
\begin{equation}
p_i^1 = \frac{p_i e^{-\lambda_i I_1}}{\sum\limits_{k=1}^{K} p_k e^{-\lambda_k I_1}}, \quad i=1,2,..., K.
\end{equation}
In general, the phase probabilities at $j^{th}$ sensing instant $T_j$ are given by
\begin{equation}
p_i^j = \frac{p_i e^{-\lambda_i (I_1 + I_2 + ... + I_j)}}{\sum\limits_{k=1}^{K} p_k e^{-\lambda_k (I_1 + I_2 + ... + I_j)}}, \quad i=1,2,..., K.  \nonumber
\end{equation}
which can be rewritten as
\begin{equation}
p_i^j = \frac{p_i^{j-1} e^{-\lambda_i I_j}}{\sum\limits_{k=1}^{K} p_k^{j-1} e^{-\lambda_k I_j}}, \quad i=1,2,..., K. 
\label{eq:pij}
\end{equation}
Let $C_S$ denote the cost per channel sensing.  Let $C_I$ denote the cost which is a measure of  the interference to PU per unit time.  The costs $C_S$ and $C_I$ can be a measure of channel sensing/switching energy and SU retransmission energy, respectively. Alternatively, they can represent the time for channel sensing/switching and the time for SU retransmission. We now define an indicator random variable $1_{Interf,j}$  to represent the SU's interference to PU in $(j+1)^{th}$ sensing interval as follows:
\begin{equation}
\begin{split}
1_{Interf,j} &  = \begin{cases}
1,  & X_{j} \leq I_{j+1}\\ 
0,  &\textit{otherwise}. 
\end{cases}
\end{split}
\end{equation} 
Then, the average amount of interference to PU in $(j+1)^{th}$ sensing interval, denoted as $E[(I_{j+1}-X_j) 1_{Interf,j}  ]$, is calculated as
\begin{equation}
\begin{split}
E[&(I_{j+1}-X_j)  1_{Interf,j } ]  = \int_{x_j=0}^{I_{j+1}}(I_{j+1}-x_j)\sum_{i=1}^{K} p_i^j \lambda_i e^{-\lambda_i x_j}dx_j 
 = I_{j+1} - \sum\limits_{i=1}^{K} p_i^j \frac{1-e^{-\lambda_i I_{j+1}}}{\lambda_i}.
\end{split} \nonumber
\end{equation}

 Let $\omega$ and $\omega^c =  1 - \omega$ be the weights (importance) that we assign  to balance the number of channel sensing by SU and Interference to PU. Thus, the average cost incurred by SU at $j^{th}$ channel sensing instant for choosing next sensing interval as $I_{j+1}$ is given by
\begin{equation} \label{eq:cost_per_stage}
C_j(I_{j+1}) = \omega C_S + \omega^{c} E[(I_{j+1}-X_j) 1_{Interf, j } ] C_I.
\end{equation}

\section{ Formulation of Optimization problem} \label{sec:opt_formulation}
We formulate the problem of optimal channel sensing interval mechanism in this section.
Let $N$ be the sensing instant at which SU detects the presence of PU.  $N$ is given by $N = \min\{ n:  (\sum_{j=1}^{n} I_j )  > X \}$. Note that the sensing interval $I_{j+1}$ is chosen by the SU at $j^{th}$ sensing instant. Let $\gamma_j(I_{j+1})$ denote the probability that the residual OFF time $X_j$ is greater than $I_{j+1}$.   It is given by
\begin{equation}
\gamma_j(I_{j+1}) = Pr(X_j > I_{j+1}) = \sum\limits_{i=1}^{K} p_i^j e^{-\lambda_i I_{j+1}}. \nonumber
\end{equation}
By using the average cost function per sensing instant given by (\ref{eq:cost_per_stage}), the average total cost  $C_{Total}$ incurred by the SU during  OFF time of the PU can be calculated as
\begin{equation}
\begin{split}
C_{Total} &= C_0(I_1) + \gamma_0(I_1) \{ C_1(I_2) + \gamma_1(I_2) \{  C_2(I_3) + \gamma_2(I_3)\{..... C_{N-1}(I_{N})\}\}\},
\end{split}
\end{equation}
The total cost function $C_{Total}$ can be rewritten as
\begin{equation}
\begin{split}
C_{Total}(I_1, I_2,..., I_N) & = C_0(I_1) + \gamma_0(I_1)C_1(I_2) + \gamma_0(I_1) \gamma_1(I_1)C_2(I_3)  + ... +  ( \prod_{i=0}^{N-2}\gamma_{i}(I_{i+1})) C_{N-1}(I_{N}),
\\& =  C_0(I_1) + \gamma_0(I_1)C_1(I_2)  + \gamma_0(I_1 + I_2)C_2(I_3)   + ... +   \gamma_{0}(\sum_{i=0}^{N-2}I_{i+1}) C_{N-1}(I_{N}),
\end{split}
\label{eq:total_cost1}
\end{equation}
where we use the fact that $\prod_{i=0}^k\gamma_i(I_{i+1}) = \gamma_0(\sum_{i=0}^{k}I_{i+1})$ to get the second equality.
By observing the SU channel sensing activity in  Fig.~\ref{fig:activity}, we can also write total cost function, $C_{Total}$ as 
\begin{equation}
C_{Total}(I_1,I_2,...,I_N) =  \omega E[N]C_{S} + \omega^{c} E[(I_1+I_2+...+I_N) - X]C_I
\label{eq:total_cost}
\end{equation}

We now formally state our optimization problem as follows. Our objective is to find the optimal channel sensing intervals $\{I^*_{j}\}_{j=1}^{N}$ such that 
$C_{Total}$ is minimized, i.e.,
\begin{equation}
\{I_1^*, I_2^*,...,I_N^*\} = \arg \min_{\{I_1,I_2,...,I_N\}}  C_{Total}(I_1,I_2,..., I_N)
\label{eq:obj_fn}
\end{equation}
Note that the variables in above optimization problem $\{I_j\}$ take values from $R^+$ and the sensing index $N$ is a function of $\{I_j\}$ and  $X$.

%\begin{figure*} 
%\centering
%\begin{subfigure}[b]{0.3\textwidth}
%\centering
%%\includegraphics[width=\textwidth]{Ns.eps}
%\includegraphics[width=\textwidth]{Ns.eps}
%\caption{\small Average number of sensing, $ {E}(N)$.}
%\end{subfigure} 
%\hfill
%\begin{subfigure}[b]{0.3\textwidth}
%\centering
%%\includegraphics[width=\textwidth]{IntPU.eps}
%\includegraphics[width=\textwidth]{IntPU.eps}
%\caption{\small Average Interference to PU.}
%\end{subfigure} 
%\hfill
%\begin{subfigure}[b]{0.3\textwidth}
%\centering
%\includegraphics[width=\textwidth]{Tcost.eps}
%\caption{\small Average Total cost, $C_{Total}$ .}
%\end{subfigure} 
%\caption{\small The performance of different sub-optimal policies against $\omega$ for light-traffic channels with HED idle times \cite{stabellini2010quantifying}: (a) average number of sensing, (b) Interference to PU (c) Total cost function $C_{Total}$ with $C_s=1$ and $C_I = 1$. }
%\label{fig:plots}
%\end{figure*}

%\begin{figure*}
%\centering
%\begin{subfigure}[b]{0.3\textwidth}
%\centering
%\includegraphics[width=\textwidth]{pictures/Ns_log16_3.eps}
%\caption{\small Average number of sensing, $ {E}(N)$.}
%\end{subfigure}
%\hfill
%\begin{subfigure}[b]{0.3\textwidth}
%\centering
%\includegraphics[width=\textwidth]{pictures/IntPU_log16_3.eps}
%\caption{\small Average Interference to PU.}
%\end{subfigure}
%\hfill
%\begin{subfigure}[b]{0.3\textwidth}
%\centering
%\includegraphics[width=\textwidth]{pictures/Tcost_log16_3.eps}
%\caption{\small Average Total cost, $C_{Total}$.}
%\end{subfigure}
%\caption{\small }
%\end{figure*}
\section{Dynamic Programming framework}
\label{sec:MDP}
A deep look into the system model and the objective function (Equations (\ref{eq:total_cost1}) and~(\ref{eq:total_cost})) suggests that the SU at each sensing instant $T_j$ has to select the next optimal sensing interval $I_{j+1}$ considering the past sensing intervals to minimize the over-all cost of this sensing process. This observation suggests \textit{Stochastic Dynamic Programming} (SDP) as a tool to solve the above optimization problem \cite{Lecture}. SDP is a  generic  method to solve very complex problems by breaking them into subproblems. In order to solve  the optimization problem using SDP, the optimal solution should be decomposable into sub-problems. The total cost function $C_{Total}$ defined in (\ref{eq:total_cost1}) clearly has a decomposable optimal structure and hence we can use SDP to arrive at the optimal solution. We formulate the SDP as
\begin{equation}
V_{j}^*(T_j) = \min_{I_{j+1}\geq 0} \{ C_{j}(I_{j+1}) + \gamma_{j}(I_{j+1}) V_{j+1}^*(T_{j+1})\},
\label{eq:DP}
\end{equation}
where $V_{j}^{*}(T_j)$ is the minimum cost at time $T_j$. If the value of $V_{N}^*(T_N) = 0$, the minimal cost  $V_{0}^{*}(T_0)$ will be same as optimal total cost $C_{Total}^*$ through recursion of (\ref{eq:DP}). %The channel sensing error can be included in the above optimization problem by replacing $\gamma_{j}(I_{j+1})$ with $\gamma_{j}^{'}(I_{j+1}) = (1-P_f)\gamma_{j}(I_{j+1})$, where $P_f$ is the probability of false alarm \cite{Liang2008}.

\subsection{Structure of Markov Decision Process}
We model the above optimization problem as a Markov Decision Process (MDP) with the state space as the set of all possible probability vectors $\textbf{P}=[p_1,  p_2, ..., p_K]$ such that $\sum_{i=1}^{K} p_i = 1$ and $p_i \geq 0$. The action space of MDP be the whole non-negative real line $R^+$. %Let the stage of MDP be the channel sensing instant of SU. 
At $j^{th}$ sensing instant, the probability vector $P^j$ is given as $P^{j} = [p_1^j,  p_2^j, ..., p_K^j]$ with phase probabilities $\{p_i^j\}_{i=1}^{K}$ of residual OFF time. %(HED with parameters $\{\lambda_i\}_{i=1}^{K}$ and $\{p_i^j\}_{i=1}^{K}$ given by Eqn~\ref{eq:pvar}) .
%\begin{equation} 
%p_i^j = \frac{p_i^{j-1} e^{-\lambda_i I_j}}{\sum\limits_{k=1}^{K} p_k^{j-1} e^{-\lambda_k I_j}}, \quad i=1,2,..., K.  \label{eq:pji}
%\end{equation}
The probability vector at zeroth sensing instant, $P^0 = [p_1, p_2,..., p_K]$,  be the initial state of MDP. The SU choose an action $I_1$ from $ R^+$ at state $P^{0}$ which will cost $C_0(I_1)$. In the first channel sensing instant, the state of the system will be in $P^1$. The SU choose an action $I_2$ which move the system to state $P^2$. Similarly, for the $j^{th}$ channel sensing instant, the state of the system will be in $P^j$ which depends only on the previous state $P^{j-1}$ and the action $I_{j}$ ( satisfies Markovian property). The cost for choosing an action $I_{j+1}$ at $P^j$ state will be $C_j(I_{j+1})$ given by (\ref{eq:cost_per_stage}).

\subsubsection{Countable state space}
Note that when a SU starts from state $P^0$, there is a countable set of states that SU can reach in future  \cite{azad2011optimal}. Since the state space is a countable set, we restrict to MDP policies that choose non-randomized action $I_{j}$ at each state and the action depends only on current state \cite{feinberg1992stationary}. 

\subsubsection{Compact action space}
The action space of the above MDP can be restricted to the compact set $[0,\overline{I}]$ without loss of optimality. The $\overline{I}$  is the upper bound on channel sensing intervals $\{I_j\}$ that the SU can choose from $ R^+$ and is given as (Lemma III.4 in \cite{azad2011optimal})
\begin{equation} \label{eq:upper_bound}
\overline{I} = (\frac{1}{\omega^c})\{\overline{v} + 1 + \frac{1}{( \min_j \lambda_j)} \}, \nonumber
\end{equation}
where  $\overline{v}$ is the upper bound on the total expected cost when SU always take channel sensing interval of unit length, $ \overline{v} := \omega^c + \omega (1+ 1/{ \min_j \lambda_j} ) C_I. $

We have shown that the MDP structure of our optimization problem has countable state space, compact action space and a non-negative cost function. From the above discussion, we conclude that the optimal policies for MDP can be restricted to non-randomized decision policies and the action space is restricted to $[0,\overline{I}]$.  Thus, the minimum total cost function can be achieved by the minimal solution of the following SDP \cite{puterman2014markov}:
\begin{equation}
V(P) = \min_{I \geq 0} \{\omega C_S + \omega^{c}  E[(I-X)  1_{ X \leq I } ] C_I  + Pr( X > I ) V(P^1) \},
\label{eq:V(P)}
\end{equation}
where the random variable $X$ follows HED with parameters $\{\lambda_j\}_{j=1}^{K}$ and phase-probabilities $P$, the probability vector $P^1$ is a function of $P$ and action $I$.
The optimal channel sensing interval $I^*$ for a given $P$ is the one that minimizes the above equation. 

\subsection{Periodic sensing interval for exponential OFF times}

%We will use the above MDP framework to derive optimal sensing intervals for channels with exponential OFF times. 
We now demonstrate the correctness of MDP framework by deriving the optimal sensing interval policy for the well-known case of channels with exponential OFF times. The exponential distribution can be considered as a special case of HED with number of phases $K=1$ with $p_1=1$ and $\lambda_1 = \lambda$. Thus the above MDP framework can be used to derive the optimal sensing interval policy for exponential PU OFF time distribution. In this case, the residual OFF time at every sensing instant has same exponential distribution and thus results in one-state MDP problem, i.e. $P^1 = P$. The optimal sensing interval is found by minimizing the total cost function of SDP given below: %in Equation~\ref{eq:V(P)}.
\begin{equation}
V(P) = \min_{I \geq 0} \{\omega C_S + \omega^{c}  E[(I-X)  1_{ X \leq I } ] C_I  + Pr( X > I ) V(P) \}.
\end{equation}
By substituting $Pr(X>I) = e^{-\lambda I}$ and
$E[(I-X)1_{X \leq I}] = I - \frac{1-e^{-\lambda I}}{\lambda} $ in the above equation and re-arranging, 
\begin{equation}
V(P) = \frac{\omega C_S + \omega^c C_I\Big\{I - \frac{1-e^{-\lambda I}}{\lambda} \Big\}}{1-e^{-\lambda I}} \label{eq:VP22}
\end{equation}

The second derivative of above function is given as
\begin{equation}
V^{''} = \frac{\omega^c C_I \lambda e^{-\lambda I}} {(1-e^{-\lambda I})^3} \Big\{ (1+e^{-\lambda I})(2 + \frac{\lambda \omega C_S}{\omega^c C_I} + \lambda I)-4\Big\}, 
\nonumber
\end{equation}
where $(1+e^{-\lambda I})(2 + \frac{\lambda \omega C_S}{\omega^c C_I} + \lambda I) > 4$ for $I \geq 0$. Thus, $V^{''} > 0$ for $I \in [0,\overline{I}]$ and hence the total cost function  $V(P)$ of SDP is a convex function.
On differentiating Eq.~(\ref{eq:VP22}) w.r.t $I$ and equating to zero, we get
\begin{equation}
\begin{split}
V^{'} & = \omega^c C_I \Bigg\{ \frac{1-e^{-\lambda I^*}(1+\frac{\lambda \omega C_S}{\omega^c C_I} + \lambda I^*) }{(1-e^{-\lambda I^*})^2}\Bigg\} = 0
\\& => 1= e^{-\lambda I^*}(1+\frac{\lambda \omega C_S}{\omega^c C_I} + \lambda I^*) \nonumber
\end{split}
\end{equation}
Multiplying both sides of above equation by $ -e^{-1-\frac{\lambda \omega C_S}{\omega^c C_I}}$,
we get
\begin{equation}
-e^{-1-\frac{\lambda \omega C_S}{\omega^c C_I}} =  e^{-1-\frac{\lambda \omega C_S}{\omega^c C_I}-\lambda I^*}(-1-\frac{\lambda \omega C_S}{\omega^c C_I} - \lambda I^*) \nonumber
\end{equation}
which can be written in the form $z = x e^x$, 
where $x=(-1-\frac{\lambda \omega C_S}{\omega^c C_I} - \lambda I^*)$ and $z= -e^{-1-\frac{\lambda \omega C_S}{\omega^c C_I}}$.
The solution $x$ of the above form $z=xe^x$ is Lambert-W function \cite{lambertW} at point $z$, i.e, 
\begin{equation}
-1-\frac{\lambda \omega C_S}{\omega^c C_I} - \lambda I^* = W_{-1}\Big (-e^{-1-\frac{\lambda \omega C_S}{\omega^c C_I}}\Big ), \nonumber
\end{equation} 
where $W_{-1}$ denotes the branch of the Lambert-W function that is real-valued on the interval $[-e^{-1},0]$ with values below -1. From the above equation, we will get the optimal sensing interval $I^*$ which is used by SU at all sensing instants (i.e. periodic sensing interval $I^*$) as
\begin{equation} \label{eq:periodic}
I^* = -\frac{1}{\lambda}  - (\omega /\omega^c) \frac{C_S}{C_I}  - \frac{W_{-1}(-e^{ - 1  - \lambda (\omega /\omega^c)*( C_S/ C_I)})}{\lambda}.
\end{equation}
Thus, we have shown that (\ref{eq:periodic}) is equivalent to the results derived in~\cite{pei2007}. We have also proven that the optimal sensing policy for exponential OFF time distribution  is periodic sensing with sensing interval $I^*$.

\section{Sub-optimal Policies}
\label{sec:suboptimal}
The Hyper-exponential distribution given in (\ref{eq:fx}) is a convex combination of exponential distributions. Using the concepts of reliability theory, we can show that HED has Decreasing Failure Rate, i.e. the probability $\gamma_j(I_{j+1})=Pr(X_j>I_{j+1})$ is increasing with increase in 'j' \cite{Shabara2015}. As a result, the optimal policy for (\ref{eq:DP}) should  account for infinite number of optimal actions/sensing intervals ${I_j^*}_{j=1}^N$ for $N \to \infty$. Moreover, the formulated MDP problem has continuous state space and action space. Thus the derivation of optimal sensing intervals (i.e actions) for HED OFF time is computationally complex and we are going for suboptimal policies. First, we adapt some of the existing sub-optimal policies \cite{azad2011optimal} for our cost function given in (\ref{eq:Ctotal}). We then suggest a new policy called "Multishot sensing interval policy" which outperforms existing sub-optimal policies in many scenarios.
\subsection{Exponential sensing interval  policy}
\begin{figure}[ht]
\centering
\includegraphics[scale=0.6, trim=0mm 0mm 0mm 0mm,clip=true]{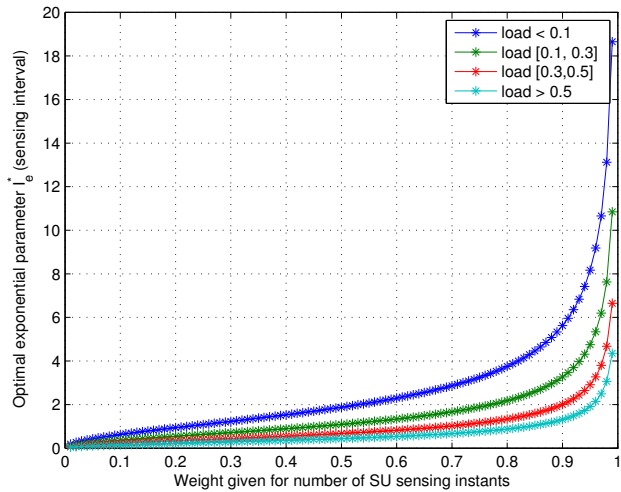}
\caption{\small Optimal exponential parameter $I_e^*$ for different weightage $\omega$ with costs $C_s=10$ and $C_I = 5$ for HED parameters in \cite{stabellini2010quantifying}. }
\label{fig:Optimal_Ie}
\end{figure}

In exponential channel sensing interval policy, the secondary user at each sensing instant selects the next sensing interval which is a realization of exponential random variable with parameter $\lambda_e$. At every sensing instant (state), the SU's sensing interval (action) is an exponential random variable with parameter $\lambda_e$.  The optimal exponential parameter $\lambda^*_e$ is derived as follows: The total cost function, $C_{Total}$ for the exponential sensing interval policy is calculated as:

\begin{equation}
\begin{split}
C_{Total}(I_1,I_2,...,I_N) & =  \omega C_{S} E[N]  + \omega^{c} E[(I_1+I_2+...+I_N) - X]C_I, 
\end{split} \label{eq:Ctotal}
\end{equation}
where $E[X]$, $E[N]$ \& $E[I_1+I_2+...+I_N]$ are 
\begin{equation}
\begin{split}
& E[X] = \sum_{j=1}^{K}p_j/\lambda_j, 
\\& E[N] = \lambda_e E[X]+ 1, 
\\& E[I_1+I_2+...+I_N] = \frac{1}{\lambda_e} E[N]  =  E[X] + \frac{1}{\lambda_e}.
\end{split}
\nonumber
\end{equation} 
Substituting the above values in (\ref{eq:Ctotal}), we will get $C_{Total}$ for exponential sensing interval policy as
\begin{equation}
C_{Total,Exp} = \omega C_s \lambda_e E(X) + \omega C_s + \frac{\omega^c}{\lambda_e}C_I 
\end{equation} 
Taking the first and second order derivative of $C_{Total,Exp}$ with respect to $\lambda_e$, we get
\begin{equation} \label{opt_lam}
\lambda_e^* = \sqrt{\frac{\omega^c C_I }{\omega C_s E[X]}} ,
\end{equation}
and the minimal total cost as
\begin{equation} 
C_{Total,Exp}^* = \omega C_s + 2\sqrt{\omega \omega^c C_s E[X] C_I}. \label{eq:C*}
\end{equation}
Thus, the SU at each sensing instant will take value from exponential distribution with parameter $\lambda_e^*$ as the next channel sensing interval. We plot the optimal parameter $\lambda_e^*$ given in (\ref{opt_lam}) against weight $\omega$ in Fig.~\ref{fig:Optimal_Ie} using numerical computation in C++. In our simulation, we vary the channel load by only varying the HED OFF times as in \cite{stabellini2010quantifying}. Thus the average PU OFF time $E(X)$ decreases with increase in channel load. We can observe from Fig.~\ref{fig:Optimal_Ie} that the mean optimal sensing interval $I_e^* = 1/\lambda_{e}^{*}$ decreases with increase in channel load condition. Further, we can also notice that SU sense the channel frequently when more importance is given to reduce interference to PU, i.e. $\omega$ $\sim$ 0.%The HED parameters given in \cite{stabellini2010quantifying} is used to plot Fig.~\ref{fig:Optimal_Ie}.

%\textbf{MATLAB simulation of SU using the derived exponential policy -- simulation running for plots}
%\begin{minipage}[b]{0.48\linewidth}
%\centering
%\includegraphics[scale=0.5, trim=7mm 0mm 0mm 7mm,clip=true]{chap_MDP/EN.eps}
%\caption{\small Average number of sensing for optimal exponential policy vs $\omega$ with costs $C_s=10$ and $C_I = 5$ for HED parameters in \cite{stabellini2010quantifying}.}
%\end{minipage}
%\hfill
%\begin{minipage}[b]{0.5\linewidth}
%\centering
%\includegraphics[scale=0.45, trim=7mm 0mm 0mm 7mm,clip=true] {pictures/Int_PU.eps}
%\caption{\small Average .}
%\label{fig:Int_PU}
%\end{minipage}

\subsection{One-stage sensing interval policy}
One-stage sensing interval policy is a policy improvement over first stage (zeroth sensing instant) of existing exponential sensing interval policy. In one-stage sensing policy, the SU uses the SDP formulation given in (\ref{eq:DP}) to select only the first sensing interval $I_1$. Thereafter, the SU follows exponential sensing interval policy by replacing random variable $X$ with $X_1$ following residual HED OFF time distribution with phase probabilities $P^1$.  The value of $V_1^*(T_1)$ in (\ref{eq:DP}) will be $C_{Total,Exp}^*(X_1)$  , as given in (\ref{eq:C*}).

%\subsection{Numerical evaluation of sub-optimal parameters}
%The optimal parameter $\lambda_e^*$ of exponential sensing interval policy is calculated using (\ref{opt_lam}). 
The optimal parameters $I_1^*$ and $\lambda_{1e}^*$ for one-stage sensing interval policy are derived as follows: \newline 
(i) Evaluate the upper bound on sensing interval, i.e. $\overline{I}$ using (\ref{eq:upper_bound}). \vspace{0.25cm} \newline 
(ii) Vary the values of $I_1$ from zero to $\overline{I}$ in steps of $\Delta$ (In our simulation, we set $\Delta = 1e^{-4}$  based on analysis of $E(X)$ ) \vspace{0.25cm} \newline 
(iii) For each values of $I_1$, calculate the cost $C_0(I_1)$ and probability vector $P^1$ using (\ref{eq:cost_per_stage}) \& (\ref{eq:pij}), respectively. For the remaining stages, the exponential sensing interval policy is used.\vspace{0.25cm} \newline
(iv) For each value of $I_1$, the parameter $\lambda_{1e}^*$ of exponential sensing interval policy is calculated using (\ref{opt_lam}) by replacing $E(X)$ with $E(X_1) = \sum_{i=1}^{K}p_i^1 \lambda_i $. Similarly, the $C_{Exp}^*$ is calculated using (\ref{eq:C*}) with $E(X_1)$. \vspace{0.25cm} \newline
(v) The total cost of one-stage policy, $C_{Total}(I_1,\lambda_{1e})$, is given as
\begin{equation} \label{eq:complex}
C_{Total,One-stage}(I_1,\lambda_{1e}) = C_0(I_1) + Pr(X > I_1)C_{Exp}^*(I_1,\lambda_{1e}) 
\end{equation} 

The value of $I_1$ which minimizes the $C_{Total,One-stage}(I_1,\lambda_{1e})$ is taken as the optimal first sensing interval $I_1^*$ and its corresponding exponential parameter is taken as $\lambda_{1e}^*$ for one-stage sensing interval policy.

%\subsection{Two-stage sensing interval policy}
%The secondary user uses DP formulation in Eq.~(\ref{eq:DP}) to select the first and second sensing intervals $I_1$ \& $I_2$ for two-stage sensing policy, respectively. Thereafter, the SU follows exponential sensing interval policy by replacing the random variable $X$ with the residual HED OFF time distribution $X_2$ with phase probabilities $P^2$. The Note that the evaluation of two stage sensing interval policy is computationally intensive but it has better performance than one-stage interval policy.

\subsection{Multishot sensing interval policy}

We propose a new sub-optimal policy called ``Multishot sensing interval policy" based on the observation that the probability vector $P^j \to [1,0,...,0]$ as $j \to \infty $ when we rearrange HED parameters such that $\lambda_1 < \lambda_2 < .... < \lambda_K$.  At zeroth sensing instant, SU assumes idle time to follow exponential random variable with parameter $\lambda_K$ and uses periodic sensing interval policy to derive  the first sensing interval $I_1$. If the channel is still idle at the first sensing instant, SU assumes that idle time was generated by exponential random variable with parameter $\lambda_{K-1}$ and uses periodic sensing interval policy to derive  $I_2$. SU keeps on changing parameter of exponential RV till it reaches $(K-1)^{th}$ sensing instant where it uses $\lambda_1$  to derive $I_{K}$. If the channel is still free from PU, the SU uses $I_{K}$ as the sensing interval for the remaining sensing instants. We will prove through simulation that multishot policy outperforms existing sub-optimal policies in most of the test-cases.

\subsection{Computational complexity of suboptimal policies}
We now compare the computational complexity in obtaining the optimal parameters of one-stage and multishot sub-optimal channel sensing policies. In case of exponential and multishot policy, the expected number of times the sensing interval is computed is a bounded constant and hence the order of complexity  is  $\mathcal{O}(1)$. For the case of one-stage sub-optimal policy, we fix the appropriate time step $\Delta$ and then evaluate Eq.(\ref{eq:complex}) for $ Z = \frac{\overline{I}}{\Delta}$ number of times to get the optimal parameters. We can observe that the order of complexity of one-stage sub-optimal policy is $\mathcal{O}(Z)$.

In similar lines to that of  one-stage sub-optimal policy, it is possible to derive Mth stage suboptimal policy but the computational complexity will be of order $\mathcal{O}({Z^M})$. As the number of stages M increases, the suboptimal policy gets closer to the optimal solution. When $M \to \infty$, we will get the optimal policy (based on the observation that the vector  $P^j \to [1 0 0 … 0]$ as $j \to \infty$ )

\section{Simulation Results} \label{sec:simulation}

We calculate the optimal parameters of different sub-optimal policies numerically using C++. Then, we simulate the PU channel occupancy patterns where OFF times are generated with HED distribution given in \cite{stabellini2010quantifying}, \cite{sharma2011comprehensive}. When the channel become free from PU, the SU will access the channel using one of the sub-optimal policies. We evaluate the performance of different sub-optimal policies in terms of total cost $C_{Total}$ through a simulator written in C++.

%For Two-stage sensing interval policy, the optimal parameters $I_1^*$, $I_2^* $ and $\lambda_{2e}^*$ are found in similar manner as that of One-stage policy. We have to find the optimal parameters which minimizes the total cost, $C_{Total,Two-stage}(I_1,I_2,\lambda_{2e})$, by varying both $I_1$ and $I_2$  for first two sensing instant and then following exponential sensing interval policy for remaining stages.
%\begin{equation}
%C_{Total,Two-stage}(I_1,I_2,\lambda_{2e}) = C_0(I_1) + Pr(X > I_1)\{ C_1(I_2) + Pr(X_1 > I_2) C_{Exp}^*(I_1+I_2,\lambda_{2e}) \} \nonumber
%\end{equation}

\begin{figure*}[ht]
\centering
\subfloat[Light traffic channel \cite{stabellini2010quantifying} ]
{{
\includegraphics[width=0.48\textwidth]{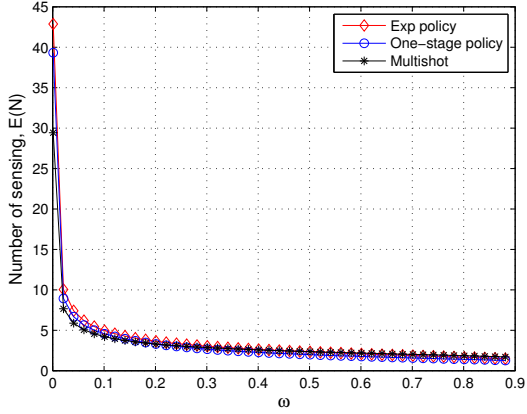}
}}
\hfill
\subfloat[Medium traffic channel \cite{stabellini2010quantifying}]
{{
\includegraphics[width=0.48\textwidth]{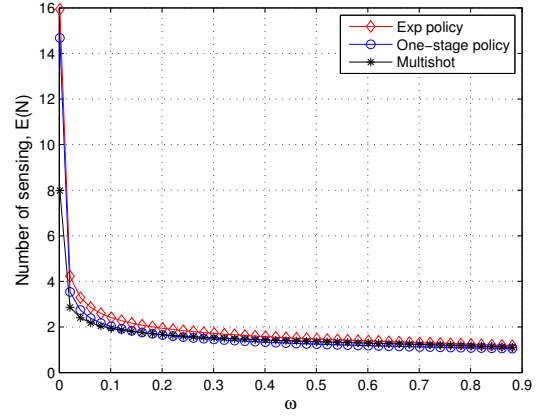}
}}
\caption{\small The average number of channel sensing $E[N]$ for different sub-optimal policies against $\omega$ with costs $C_s=1$ and $C_I = 1$. }
\label{fig:plot1}
\end{figure*}
\begin{figure*}[ht]
\centering
\subfloat[Light traffic channel \cite{stabellini2010quantifying} ]
{{
\includegraphics[width=0.48\textwidth]{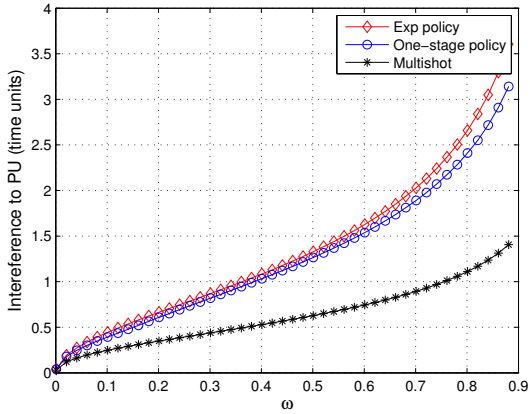}
}}
\hfill
\subfloat[Medium traffic channel \cite{stabellini2010quantifying}]
{{
\includegraphics[width=0.48\textwidth]{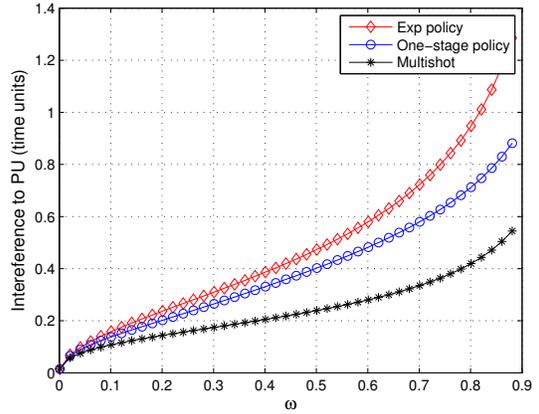}
}}
\caption{\small  The average interference to PU (in time units) for different sub-optimal policies against $\omega$ with costs $C_s=1$ and $C_I = 1$.  }
\label{fig:plot2}
\end{figure*}
\begin{figure*}[ht]
\centering
\subfloat[Light traffic channel ]
{{
\includegraphics[width=0.48\textwidth]{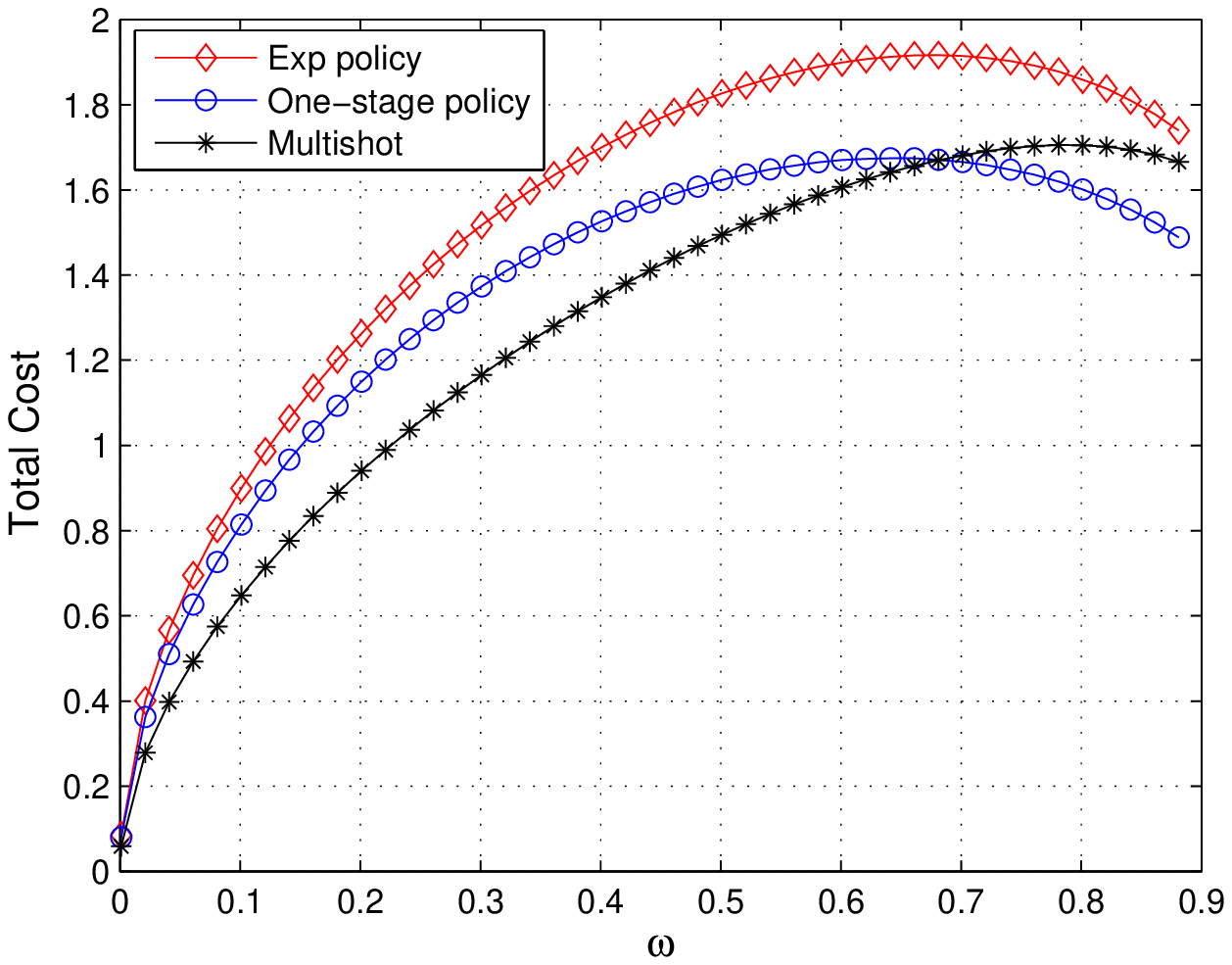}
}}
\hfill
\subfloat[Medium traffic channel ]
{{
\includegraphics[width=0.48\textwidth]{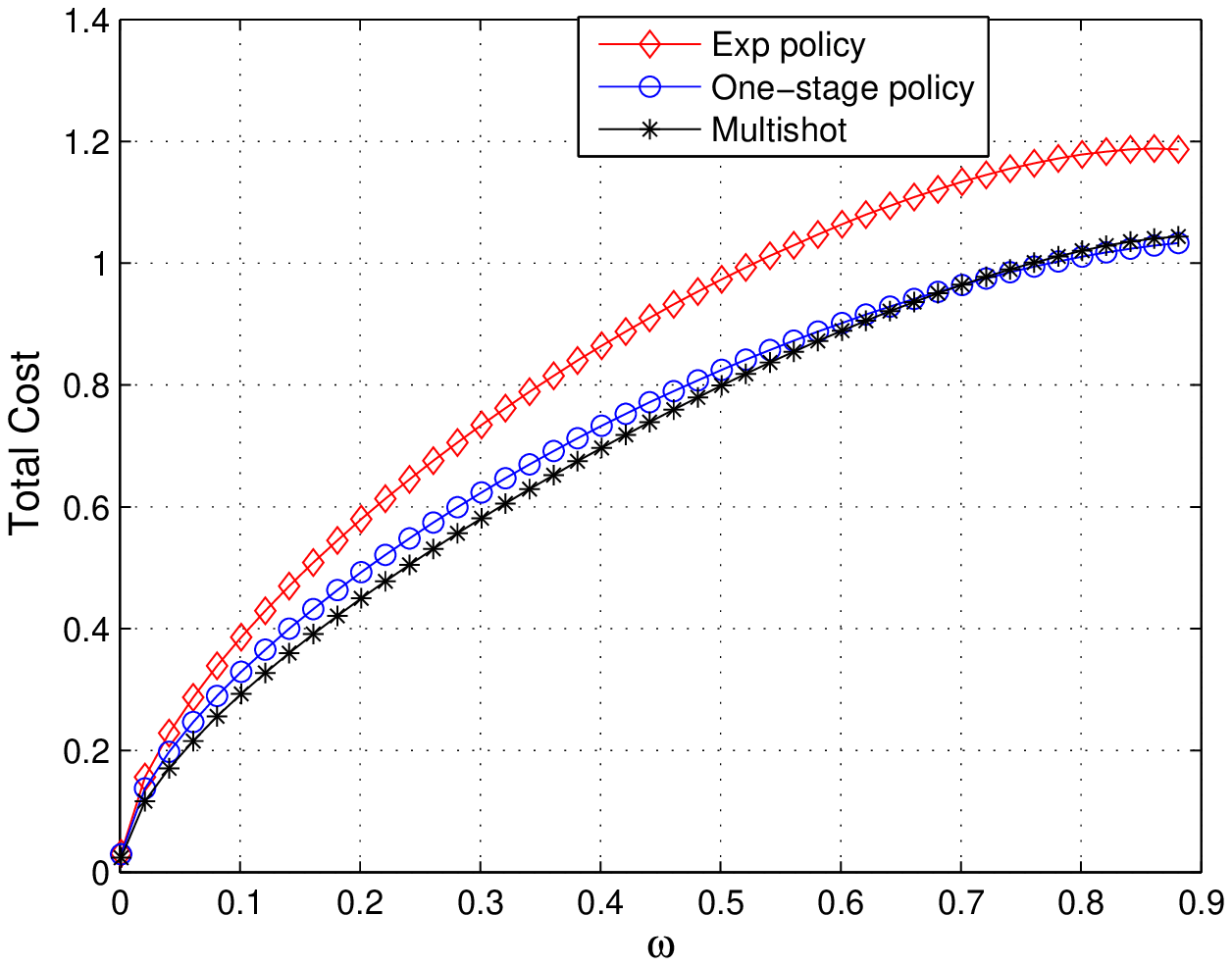}
}}
\caption{\small The average total cost $C_{Total}$ for different sub-optimal policies against $\omega$ with costs $C_s=1$ and $C_I = 1$.  }
\label{fig:plot3}
\end{figure*}

The optimal sensing intervals of multishot sub-optimal policy, $\{I^*_1, I^*_2,..., I^*_K\}$, are calculated using~(\ref{eq:periodic}) for parameters $\{ \lambda_K, \lambda_{K-1}, ..., \lambda_1\}$, respectively. From $(K+1)^{th}$ sensing instant, the SU always choose $I^*_K$ as the channel sensing interval.

In our simulation, we generated the channel occupancy model using two sets of HED parameters given in \cite{stabellini2010quantifying} and \cite{sharma2011comprehensive} (Light traffic -- load $<$ 0.1 and Medium traffic -- load $\in$ [0.3,0.5]). Then, we evaluated the performance of different sub-optimal policies in terms of average number of channel sensing $E[N]$, average interference to PU (in time units) and $C^*_{Total}$ which are plotted in Figs.~\ref{fig:plot1} -- \ref{fig:plot3}, respectively. We observed that the interference to PU is less in case of multishot policy as compared to other sub-optimal policies.

The performance of all sub-optimal policies for channels with cost functions $C_S=5$ and $C_I=1$ are tabulated in Table~\ref{tab:performance}. We can observe that multishot policy outperforms exponential sensing interval policy in terms of total cost $C_{Total}$ in all type of traffic conditions.  The performance of sub-optimal policies also depends on channel's traffic conditions as well as costs $C_S$ and $C_I$.  For example, we observe from Fig.~\ref{fig:plot3}(a) and Table~\ref{tab:performance} that the cross-over point of $C_{Total}$ for multishot and one-stage sub-optimal policies varies with change in costs. 

In general, the proposed multishot policy outperforms one-stage sub-optimal policy when more weightage is given to reduce interference to PU. One-stage policy outperforms multishot policy if we give more importance to reduce the number of channel sensing \footnote{The cross-over point of $C_{Total}$  for multishot and one-stage policy varies with respect to $C_S$, $C_I$ and also HED parameters.}.  However, the major advantage of using multishot  sub-optimal policy is that the complexity in calculating the parameters of one-stage policy is very high ($\mathcal{O}(Z)$) as compared to multishot policy  ($\mathcal{O}(1)$).

%\begin{figure*} [ht]
%\centering
%\subfloat[Average number of sensing, $ {E}(N)$]
%{{
%\includegraphics[width=0.3\textwidth]{chap_MDP/Ns_log14.eps}
%}}
%\hfill
%\subfloat[Average Interference to PU ]
%{{
%\includegraphics[width=0.3\textwidth]{chap_MDP/IntPU_log14.eps}
%}}
%\hfill
%\subfloat[Average Total cost, $C_{Total}$]
%{{
%\includegraphics[width=0.3\textwidth]{chap_MDP/Tcost_log14.eps}
%}}
%\caption{\small The performance of different sub-optimal policies against $\omega$ for channels with 5-phase HED idle times \cite{stabellini2010quantifying} with costs $C_S=1$ and $C_I = 1$. }
%\label{fig:plots}
%\end{figure*}

\begin{table*}
\tiny
\centering
\caption{Average number of channel sensing, intereference to PU and the $C_{Total}$ for costs $C_S = 5$ \& $C_I = 1$}
\begin{tabular}{c | c | c c c c  | c c c c | c c c c }
\hline
HED Params & Policy & \multicolumn{4}{|c|} {No.of.sensing, $E[N]$} &  \multicolumn{4}{|c|} {Interference} &  \multicolumn{4}{|c}{Total cost, $C_{Total}$} \\
\cline{3-14}
 & & $\omega=0.1$ & $0.3$ & $0.5$& $ 0.7$& $\omega = 0.1$ & $0.3$ & $ 0.5$& $0.7$&  $\omega = 0.1$ & $ 0.3$ & $0.5$& $ 0.7$  \\
 \hline

\multirow{3}{*} {Light Traffic \cite{stabellini2010quantifying}} 
& Exponential & 2.778 & 1.905 & 1.593 &  1.388 & 9.881x$10^{-1}$ & 1.941 & 2.963 & 4.529&    2.278 & 4.217 & 5.463 & 6.217  \\ 
& One-stage & 2.420 & 1.604 &  1.346 &  1.193 & 9.371x$10^{-1}$ & 1.814 & 2.651 & 3.812 & 2.054 & 3.675 & 4.691 &  5.320 \\
& Multishot & 2.684 & 2.048 & 1.795 & 1.604 &  4.888x$10^{-1}$ &  8.616x$10^{-1}$ & 1.209 & 1.668 &  1.782 & 3.674 & 5.092 & 6.116 \\ \hline

\multirow{3}{*} {Medium Traffic \cite{stabellini2010quantifying}} 
& Exponential & 1.634 & 1.323 & 1.211 &  1.138 & 3.521x$10^{-1}$ & 6.918x$10^{-1}$ & 1.056 &  1.613 &    1.134 & 2.468  & 3.556  & 4.468  \\ 
& One-stage & 1.381 & 1.137 & 1.072 &  1.038 & 3.017x$10^{-1}$ &  5.594x$10^{-1}$ & 7.708x$10^{-1}$ & 1.022 &    9.622x$10^{-1}$ & 2.098 & 3.065 &  3.941 \\
& Multishot & 1.469 & 1.245 & 1.148 & 1.076 &   1.915x$10^{-1}$ &  3.231x$10^{-1}$ & 4.600x$10^{-1}$ & 6.655x$10^{-1}$ &  9.068x$10^{-1}$ & 2.093 & 3.099 & 3.967 \\ \hline

 \multirow{3}{*} {5-phase HED \cite{sharma2011comprehensive}} 
& Exponential & 2.912 & 1.973 & 1.637 & 1.417 & 1.065 & 2.093 & 3.196 & 4.880 &    2.415 & 4.425 & 5.691 & 6.424  \\ 
& One-stage & 1.599 & 1.319 & 1.226&  1.167& 5.015x$10^{-1}$ & 8.035x$10^{-1}$ & 1.107 & 1.555 & 1.251 & 2.541 & 3.619 &  4.554  \\
& Multishot & 1.498 & 1.250 & 1.176 & 1.129&   5.065x$10^{-1}$ &  6.318x$10^{-1}$ & 9.506x$10^{-1}$ & 1.476 &  1.205 & 2.317 & 3.416 & 4.395 \\ \hline 
\end{tabular}
\label{tab:performance}
\end{table*}
\section{Effect of Sensing parameters and delayed occupancy} \label{sec:SensingParams}

\subsection{Effect of delayed occupancy} 

When the channel is busy due to transmission of PU, the SU has to periodically sense the channel for spectrum opportunity following a busy-period channel sensing strategy. As a result the SU cannot occupy the channel as soon as it is released by the PU resulting in the missed spectrum opportunity.   We now account for the effect of the delayed channel occupancy by the SU on the sub-optimal channel sensing policies. 

Let the random variable X denote the OFF time of the PU. Let the interval between the time the channel becomes free until it is sensed and occupied by the SU be denoted the random variable $M$. Let the \textit{p.d.f} of $M$ be denoted as $f_M(m)$. The residual channel idle time, after subtracting the missed opportunity, from the PU's OFF time still follows an HED distribution but with different phase probabilities as derived below. Then, the \textit{p.d.f} of residual channel idle time, denoted as $X^d$, is calculated as
\begin{equation}
\begin{split}
    f_{X^d}(x) &= \int_{m=0}^{\infty} f_{X|M}(m)f_M(m)dm, \\ & = \int_{m=0}^{\infty}\mathlarger{\mathlarger{\sum}}_{i=1}^K \Bigg\{ \frac{p_i e^{-\lambda_i m}}{\sum_{k=1}^{K} p_ke^{-\lambda_k m}} \lambda_i e^{-\lambda_i x} \Bigg\} f_M(m) dm
    \\& = \mathlarger{\mathlarger{\sum}}_{i=1}^{K} \lambda_i e^{-\lambda_i x} \int_{m=0}^{\infty} \frac{p_i e^{-\lambda_i m}}{\sum_{k=1}^{K} p_k e^{-\lambda_k m}} f_M(m) dm
    \\& =  \mathlarger{\mathlarger{\sum}}_{i=1}^{K} \lambda_i e^{-\lambda_i x} p_i^d \nonumber
\end{split}
\end{equation}
Thus the remaining channel idle time due to delayed occupancy $X^d$ follows HED, irrespective of SU's busy-period sensing interval mechanism, with same $\{ \lambda_i\}_{i=1}^K$ but with different phase probabilities $\{p_i^d \}_{i=1}^{K}$.   \footnote{In multi-channel scenario, the validity of assumption depends on sensing duration, channel sensing order, channel switch delay and  transmit/receive mode switch delays (for half-duplex SU).} 

\begin{figure}[ht]
\centering
\includegraphics[scale=0.6, trim=0mm 0mm 0mm 0mm,clip=true]{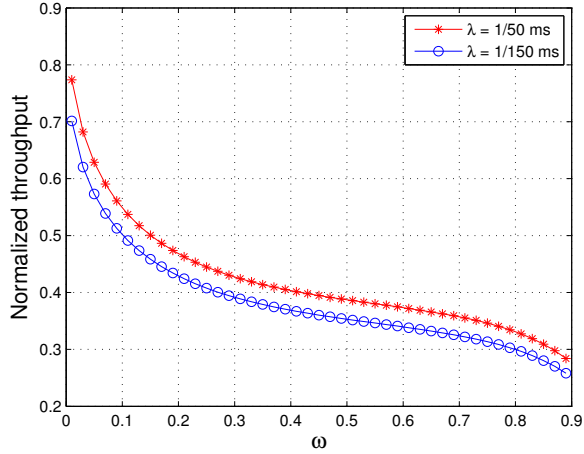}
\caption{\small Normalized throughput of secondary network for delayed occupancy (Multishot policy) with $C_S = 5$ and $C_I = 1$. HED parameters are given in \cite{stabellini2010quantifying}. }
\label{fig:Lambda_curve}
\end{figure}

For example, we have considered exponential sensing interval policy with parameter $\lambda$ for SU's busy-period sensing. As a result of memory-less property of exponential distribution, the missing opportunity due to delayed occupancy will also follows same exponential distribution, i.e. $f_M(m) = \lambda e^{-\lambda x}$. We have plotted the normalized throughput of SU against $\omega$ for different values of $\lambda$ in Fig.~\ref{fig:Lambda_curve}. The normalized throughput decreases with decrease in weightage factor for interference in total cost function. When the weightage for interference to PU decreases, we will have larger optimal sensing intervals $I_i$ and hence lesser throughput due to interference with PU.

\subsection{Effect of sensing error and sensing duration}
Two important parameters that affect the performance of channel sensing are (i) probability of detection $P_d$ and (ii) probability of false alarm $P_f$ which are defined as,
\begin{equation*}
\begin{split}
   & P_d = Pr(CH_{sensed} = Busy| CH = Busy)
   \\& P_f = Pr(CH_{sensed} = Busy| CH = Idle)
    \end{split}
\end{equation*}
The probability of false alarm $P_f$ can be expressed in terms of $P_d$, channel sensing time $T_{sense}$ and signal-to-noise ratio (SNR) $\zeta$ of complex valued PU signal as \cite{Pei2011}
\begin{equation}
P_f =Q(\sqrt{2 \zeta +  1}Q^{-1}(P_d) + \sqrt{T_{sense}f_s} \zeta) 
\label{eq:Pf}
\end{equation}
where $Q(.)$ is the tail probability of standard normal distribution, $f_s$ is the sampling frequency. The target probability of signal detection $P_d$ is usually set by regulatory bodies to avoid interference to PU. For example, IEEE 802.22 WRAN working group sets the target $P_d = 0.9$ in the worst-case scenario of $\zeta =-20$ dB. Thus with received SNR $\zeta$ and target $P_d$, we can calculate false alarm $P_f$ for different values of $T_{sense}$. 

The channel sensing error can be included in the cost function $V_{j}^{*}(T_j)$ give by equation \ref{eq:DP} of stochastic dynamic programming framework as 
\begin{equation}
V_{j}^*(T_j) = \min_{I_{j+1}\geq 0} \{ C_{j}(I_{j+1}) + \gamma_{j}(I_{j+1})(1-P_f) V_{j+1}^*(T_{j+1})\},
\label{eq:DP2}
\end{equation}
Note that the probability of detection $P_d$ and other channel sensing parameters are indirectly captured by $P_f$ as shown in (\ref{eq:Pf}). 
We have evaluated the performance of our proposed multishot  policy for different values of $P_f$, i.e. for different channel sensing duration $T_{sense}$,  for a fixed $P_d = 0.9$, $\zeta = -20$ dB, and $f_s = 20$ MHz. Whenever the channel is sensed busy (either due to PU reappearance or false alarm), SU follows busy-period sensing interval policy till the channel is sensed idle and revert back to multishot policy (restarts from $I_1^*$) after regaining the channel. In our simulation, we have assumed exponential policy with parameter $\lambda = 1/10ms$ as SU's busy-period sensing interval policy. 

\begin{figure*}[ht]
\centering
\subfloat[For fixed $\omega = 0.5$ with varying $P_f$ values]
{{
\includegraphics[width=0.48\textwidth]{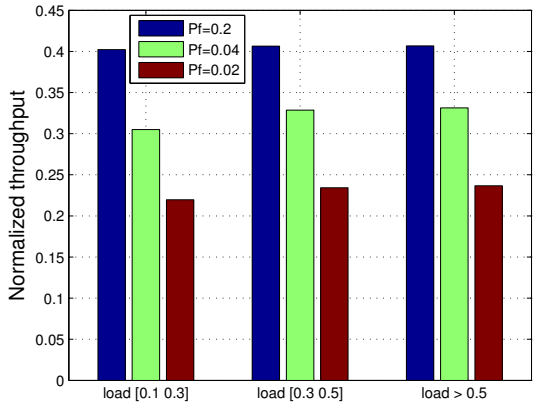}
}}
\hfill
\subfloat[ For fixed $P_f= 0.02$ ($T_{sense}$ = 60ms) with varying $\omega$ values  ]
{{
\includegraphics[width=0.48\textwidth]{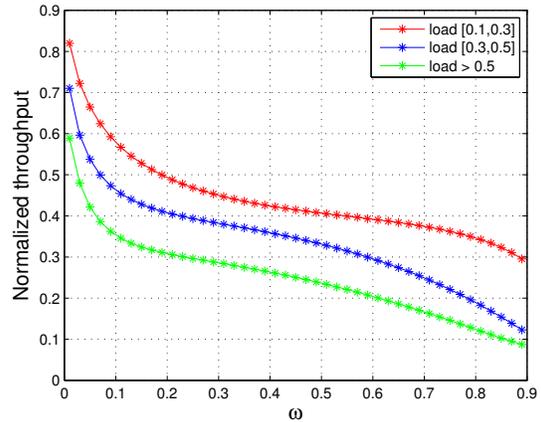}
}}
\caption{\small Normalized throughput of secondary network (Multishot policy) with $C_S = 5$ and $C_I = 1$.  }
\label{fig:Pf}
\end{figure*}

We have also incorporated channel sensing duration $T_{sense}$ which is a function of $P_f$. The normalized throughput of SU for varying channel load condition is plotted against $P_f$ in Fig.~\ref{fig:Pf}(a) for $\omega = 0.5$. We can observe that the normalized throughput decreases with increase in $P_f$. However, we didn't observe much difference in normalized throughput with respect to different channel loads. The reason being that the normalized throughput is measured as the fraction of time SU uses the channel idle time for packet transmission.  Similarly, the normalized throughput is plotted against $\omega$ for a fixed $P_f = 0.02$ in Fig.~\ref{fig:Pf}(b).  

We now discuss the effect of finite channel sensing duration on the total cost function of various sub-optimal policies. Any optimal (even sub optimal) solution would choose sensing interval $I_i$ that are much larger than the sensing duration $T_{sense}$. Else the fraction of time spent on sensing will be a large overhead. Under this condition, the sensing duration has minimal impact on our total cost. Our total cost $C_{Total}$ depends on the number of sensing made and the interference to PU. Finite sensing duration adds a small constant to the successive sensing interval chosen, and slowly drifts the sensing points as compared to the ideal case of ``Zero sensing duration".  If $N$ is the expected number of sensing done in ideal case , with finite sensing duration case it will be around "$N-1$". Therefore the error involved in total sensing  cost is just of the order $C_S$.

\section{Conclusion} \label{sec:conclusion}
In this paper, we have considered optimal channel sensing policies for channels with heavy-tailed idle time distribution, which are modeled as HED. We have shown that the periodic sensing is not optimal when channel's traffic deviates from the exponential distribution. The optimization problem, with an objective to minimize the number of SU's channel sensing and SU's interference to PU, is formulated. The structure of optimal solution  is deduced through the MDP and dynamic programming framework. By showing that the state and action space of MDP are continuous, we proposes sub-optimal channel sensing interval policy called `Multishot sensing interval policy' that minimizes the cost for sensing and interference to PU. Finally, we have compared the performance of our proposed Multishot sensing interval policy with other existing sub-optimal policies in literature for various channel traffic conditions.

%In this work, we designed optimal channel sensing policy for channels with heavy-tailed idle times, which are modeled with HED. We have shown that the periodic sensing is not optimal when channel's traffic deviates from exponential distribution. We then proposed dynamic programming framework using which one can derive optimal/sub-optimal channel sensing intervals that minimizes the cost for sensing and interference. 

%In our work, we have designed the optimal channel sensing interval framework for SUs in cognitive radio network where heavy-tailed PU OFF times are modeled with Hyper-exponential distribution. The optimization problem, with an objective to minimize the number of SU's channel sensing and SU's interference to PU, is formulated and then the structure of optimal solution  is deduced through the MDP and dynamic programming framework. Since the state and action space of MDP are continuous, we proposes sub-optimal channel sensing interval policies. We have evaluated the performance of different sub-optimal policies under various PU traffic conditions. We have shown that the optimal sensing interval policy for realistic PU channel occupancy patterns is different from that of constant sensing interval policy, which is an optimal policy for exponential OFF times.

%\bibliographystyle{IEEEtran}
%\bibliography{reference.bib}
\end{document}